\documentclass[conference]{IEEEtran}

\ifCLASSINFOpdf
 
\else

\fi

\usepackage{graphicx}
\usepackage{booktabs}
\usepackage{stfloats}
 \usepackage{amsmath} 
\usepackage{multirow}
\usepackage[table,xcdraw]{xcolor}
\usepackage{subcaption}
\usepackage{rotating}
\usepackage{tabularray}
\usepackage{hhline}
\usepackage{tabularx}

\begin{document}

\title{Color and Sentiment: A Study of Emotion-Based Color Palettes in Marketing}

% % author names and affiliations
% % use a multiple column layout for up to three different
% % affiliations
% \author{\IEEEauthorblockN{Shagyrov Maksat}
% \IEEEauthorblockA{Faculty of Information Technology\\
% Almaty, Kazakhstan \\
% Email: m\_shagyrov@kbtu.kz}
% }

\author{\IEEEauthorblockN{Maksat Shagyrov ,
Pakizar Shamoi\IEEEauthorrefmark{1}}
\IEEEauthorblockA{School of Information Technology and Engineering \\
Kazakh-British Technical University\\
Almaty, Kazakhstan\\
Email: 
\IEEEauthorrefmark{1}p.shamoi@kbtu.kz
% \IEEEauthorrefmark{2}ar\_karatayev@kbtu.kz,
% \IEEEauthorrefmark{3}an\_ogorodova@kbtu.kz,
}
}

\maketitle

\IEEEpeerreviewmaketitle

\begin{abstract}
It's widely recognized that the colors used in branding significantly impact how a brand is perceived. This research explores the influence of color in logos on consumer perception and emotional response. We investigate the associations between color usage and emotional responses in food and beverage marketing. Using a dataset of 644 companies, we analyzed the dominant colors in brand logos using k-means clustering to develop distinct color palettes. Concurrently, we extracted customer sentiments and emotions from Google Maps reviews of these companies (n=30,069), categorizing them into five primary emotions: Happiness, Anger, Sadness, Fear, and Surprise. These emotional responses were further categorized into four intensity levels: Low, Medium, Strong, and Very Strong, using a fuzzy sets approach. Our methodology involved correlating specific color palettes with the predominant emotional reactions associated with each brand. By merging the color palettes of companies that elicited similar emotional responses, we identified unique color palettes corresponding to each emotional category. Our findings suggest that among the food companies analyzed, the dominant emotion was Happiness, with no instances of Anger. The colors red and gray were prevalent across all emotional categories, indicating their importance in branding. Specific color-emotion correlations confirmed by our research include associations of yellow with Happiness, blue with Sadness, and bright colors with Surprise. This study highlights the critical role of color in shaping consumer attitudes. The study findings have practical implications for brand designers in the food industry.

% While the general role of color in marketing is well-documented, its specific impact on logo design remains underexplored. 
% Our study aims to address this gap by examining how color choices in logos affect brand perception. 
\end{abstract}


\section{Introduction}

Color plays a significant role in marketing and consumer behavior. Colors carry meaning and elicit emotions, which are important factors in establishing consumer sentiments toward a brand. Brand logos are essential for brand recognition and loyalty in the competitive food and beverage business.

% For instance, chromatic identity in branding varies across global and local markets\cite{Caivano2007}, underscoring the symbolic and cultural significance of color. Studies have shown that color preferences and meanings are context-dependent, influenced by cultural background and individual experiences\cite{Thomas}.

 Research has shown that colors can evoke emotions, influence sentiment, and impact purchasing decisions \cite{Amencherla2017Color}, \cite{Mofarah2013How}, \cite{Boshuo2023Sentiment}. Sentiment analysis of color-related content on social media can provide valuable insights for marketers \cite{Boshuo2023Sentiment}, \cite{Rambocas2018Online}. Studies have found that certain colors, such as purple and pink, are associated with more positive sentiments, while others, like brown and red, are perceived more negatively \cite{Boshuo2023Sentiment}. Color preferences can vary over time and may be influenced by factors such as texture and lighting \cite{Mofarah2013How}, \cite{Mathieu2009La}. Analyzing color-based sentiment in product reviews and images can help businesses make strategic decisions about positioning and marketing \cite{Lodha2022Sentiment}, \cite{Alasmari2020Sentimental}. The psychological impact of colors in marketing is a valuable tool for attracting consumers and differentiating products in a crowded market \cite{Hunjet2017PSYCHOLOGICAL}.

This study examines the primary colors in the logos of 644 food and beverage companies using k-means clustering. It then links these colors to fuzzy emotional reactions derived from 30,069 Google Maps reviews of these companies, using fuzzy logic to sort these emotions by intensity. This allows us to determine distinct color-emotion pairings.

The paper has the following contributions:
\begin{itemize}
    \item \textit{Empirical Analysis of Color-Emotion Associations.} By analyzing the color palettes of 644 food and beverage companies and correlating these with consumer emotional responses from online reviews, the study provides empirical evidence on how consumers perceive colors in logos. This contributes to the understanding of the psychological impact of color in branding.
    \item \textit{Application of Fuzzy Sets in Emotion Analysis. }We used fuzzy sets to assess emotional intensities and further connect and correlate emotions with dominant colors extracted using k-means clustering.
\end{itemize}

The paper is structured as follows: Section 1 is an Introduction. Section II provides an overview of related research. Next, Section III describes methods, including sentiment, emotion, and color analysis. Experimental Results are presented in Section IV. Future works and limitations are discussed in Section V. Finally, Section VI provides the concluding remarks of the study.

\section{Related Work}

The influence of color on consumer perception and emotional response has been extensively studied, with numerous researchers affirming color's powerful role in branding and marketing strategies. 

% This scientific approach to measuring color perception complements the examination of color aesthetics and context-dependency \cite{Shamoi22}, which addresses the subjective and context-bound nature of color harmony and impressions across domains like art and fashion.

% Research on the effects of color, lighting, and price fairness on purchase intentions further elucidates the complex relationship between environmental cues and consumer behavior by highlighting the mediating role of cognitive and affective associations \cite{Babin2003}. This complexity is mirrored in studies that examine the strategic use of color to influence brand perception and consumer behavior \cite{Labrecque2012}, offering empirical evidence of color's power to shape brand personality dimensions. Investigations into consumer color choices \cite{grossman1999what} and the context-dependence of color associations \cite{Labrecque2013} apply associative learning frameworks to understand how consumers develop preferences for specific colors, pointing towards the necessity of considering the specific context in which color cues are presented.

Studies on the effects of color, lighting, and price fairness elucidate the complex relationship between environmental cues and consumer behavior, emphasizing cognitive and affective associations \cite{Babin2003}. Further research explores the strategic use of color in branding \cite{Labrecque2012}, demonstrating color's influence on brand personality and consumer preferences \cite{grossman1999what, Labrecque2013}, underscoring the importance of contextual factors in color perception. Another study on measuring color perception complements the examination of color aesthetics and context-dependency \cite{Shamoi22}, highlighting the subjective nature of color harmony across various domains.

% The review of color research in marketing \cite{Amsteus2015} calls for a renewed focus on the significance of color, proposing an integrated framework that encompasses both embodied and referential meanings of color. This comprehensive perspective is crucial for navigating the complexities of color perception and its marketing implications. Studies on color coordination and the use of fuzzy sets for modeling aesthetic preferences \cite{Ghaderi2015}, \cite{fss} leverage fuzzy logic to predict preferences based on color harmony, illustrating the potential of fuzzy approaches to address the subjective and approximate nature of aesthetic judgments.

The review of color research in marketing \cite{Amsteus2015} emphasizes the need for a comprehensive framework to understand the dual meanings of color, which is essential for deciphering color perception in marketing. Research on color coordination and aesthetic preferences using fuzzy sets \cite{Ghaderi2015} shows how fuzzy logic can model and predict color preferences, demonstrating its effectiveness in handling the subjectivity of aesthetic judgments.

% The methodology for image retrieval based on fuzzy dominant colors in e-commerce \cite{Shamoi2020} and the examination of color-emotion associations in abstract art \cite{Shamoi2016} both highlight the importance of color in evoking emotional responses and its application in digital environments, suggesting ways to bridge the semantic gap between color features and high-level consumer perceptions. Research into the determinants of children's color-emotion associations \cite{JunJo} and the use of fuzzy approaches for emotion recognition in computer games for children \cite{Kozlov2023} expand the understanding of emotional responses, emphasizing the diversity of influences, including culture and nature, on color perception from an early age.

The methodologies exploring fuzzy dominant colors in e-commerce \cite{Shamoi2020} and color-emotion associations in abstract art \cite{Shamoi2016} demonstrate color's role in eliciting emotional responses and its digital applications. Further research into children's color-emotion associations \cite{JunJo} and emotion recognition in children's computer games using fuzzy approaches \cite{Kozlov2023} broadens our understanding of the varied influences on color perception from an early age, including cultural and natural factors.

Sentiment and emotion analysis have become crucial in marketing research, allowing companies to garner significant insights from online reviews and social media interactions \cite{Rambocas2018Online}, \cite{Anuj2015Mining}. This approach utilizes natural language processing and machine learning techniques to evaluate textual content methodically, offering instantaneous assessments of consumer sentiments \cite{C2020Sentiment}, \cite{S2020Insight}. Sentiment analysis provides marketers with a practical way to gather and analyze extensive data concerning consumer attitudes and perceptions of brands \cite{Rambocas2018Online}. Its use spans various marketing domains such as product feedback, influencer marketing, and brand recognition \cite{G2021Sentiment}, \cite{Priyanka2022Sentiment}. However, implementing sentiment analysis poses technical, practical, and ethical hurdles \cite{Rambocas2018Online}. Over the last twenty years, the domain has seen considerable growth, marked by greater cross-disciplinary cooperation and expanded uses in marketing communications \cite{Pablo2020Opinion}.

Research shows that emotions significantly influence purchasing choices at every stage \cite{D2010New}. To leverage this, marketers are developing platforms and frameworks for analyzing consumer emotions through various methods, including analysis of social media content \cite{Al2015Applying} and neuromarketing techniques \cite{Gill2020Review}. These approaches enable businesses to gain insights into consumers' emotional responses, leading to better marketing strategies and customer experiences \cite{Navarro2016Emotional}. Analyzing emotions in marketing contexts presents unique challenges, such as accounting for multilingual and multicultural factors in global markets \cite{Mihael2015MixedEmotions}.

Collectively, these studies paint a detailed picture of the role of color and sentiment analysis in marketing and consumer perception, suggesting a promising field for applying new methodologies uniting these methods.

\section{Methodology}

\begin{figure*}[hbt]
\centering
\includegraphics[width=\textwidth]{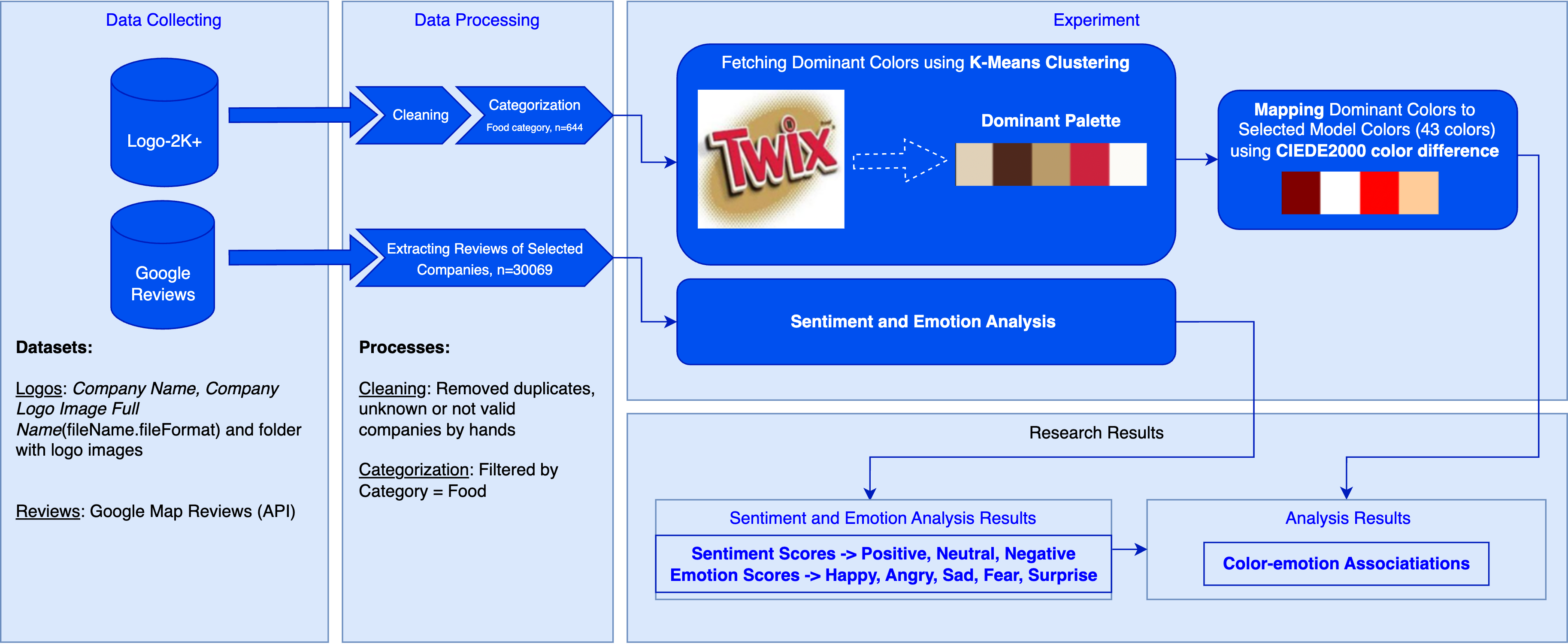}
\caption{Overview of the Methodology}\label{visina8}
\end{figure*}

\subsection{Proposed Approach}
Our research approach includes several key stages, integrating diverse data sources and analytical techniques to explore the influence of color in food company logos on consumer perception and emotional response. The methodology figure in Fig. \ref{visina8} visually represents our approach. 

Our study begins with the collection of two primary datasets: one consisting of company logos and another comprising Google Maps reviews, both extracted using specialized scripts. To evaluate consumer emotional responses to the reviews, we utilize two sentiment analysis tools: the Valence Aware Dictionary and Sentiment Reasoner (VADER) and the text2emotion model. VADER helps quantify sentiment scores that indicate the intensity of emotions, while text2emotion categorizes emotions into distinct classes, providing a deeper understanding of the emotional content in consumer feedback. For the visual aspect of our study, we apply k-means clustering to the logos to identify and quantify the dominant colors present in the logo.

\subsection{Data}
In our study, we use two datasets for logo images and company reviews.
\subsubsection{Logo Images}
The Logo-2K+ dataset \cite{logo_dataset} offers various logo classes derived from actual logo images. It consists of 167,140 images divided into ten main categories and 2,341 subcategories. The main categories include Food, Clothes, Institutions, Accessories, Transportation, Electronics, Necessities, cosmetics, Leisure, and Medical. For our study, we used logos of Food companies (n=763), but we failed to parse reviews for some of them, resulting in the total number of analyzed companies equal to 644.

\subsubsection{Company Reviews}
The company reviews dataset was collected by extracting reviews from Google Maps using Google API. Table \ref{datasettable} demonstrates the examples of reviews. The reviews extraction script automates the collection of Google Maps reviews. It begins by loading the company data and then uses the Google Maps API to search for each food company from the logo dataset and retrieve reviews. The reviews are then compiled into a structured format, including details like author, rating, and review text. The process is repeated for each company, and the results are combined into a single dataset. As a result, we have more than 30,069 reviews.
% \begin{table*}
%     \centering
%     \includegraphics[width=\linewidth]{images/SCIS/reviews_dataset.png}
%     \caption{Extracted Reviews with Score, Text}
%     \label{fig:reviews}
% \end{table*}

\begin{table*}[ht]
\centering
\caption{Examples of reviews of food companies extracted using Google API}
\begin{tabular}{|c|c|c|c|>{\raggedright\arraybackslash}p{5cm}|c|}
\hline
\textbf{ID} & \textbf{Company Name} & \textbf{Category} & \textbf{Score} & \textbf{Text} & \textbf{Time} \\ \hline
15708 & Tang & Food & 1 & Irresistible!!! The noodle box tastes... & 2023-06-15 09:36:37 \\ \hline
13572 & Domino's Pizza & Food & 1 & Stopped by while traveling to grab a fast... & 2020-08-05 18:20:53 \\ \hline
6190 & Stanley & Food & 1 & Absolute rubbish and no response... & 2022-01-17 06:25:24 \\ \hline
12249 & Shredded Oats & Food & 4 & Lovely, clean, spacious and well stocked... & 2019-10-29 17:23:04 \\ \hline
19669 & Lipton Ice Tea & Food & 5 & Dutch style, for those who like to play... & 2024-01-29 15:12:18 \\ \hline
17085 & Frosted Toast Crunch & Food & 5 & Alejandro is extremely exceptional. His... & 2021-09-27 08:22:42 \\ \hline
\end{tabular}
\label{datasettable}
\end{table*}

% Before implementing our models, we undertook a thorough cleaning of the Logos dataset. This involved removing duplicate entries for the same company and excluding logos of unknown companies. This preprocessing step refined our dataset to 229 distinct companies, each represented by its name and a vector image. To enhance the dataset, we cat- egorized the companies into relevant industry sectors (e.g., Airline, Beverages, Clothing, Fast Food) using ChatGPT.

% % \subsubsection{Data Preparation}
% For the data preparation phase, we will perform feature extraction by converting the logo images into numerical features through image processing techniques. This step is crucial for transforming visual data into a format suitable for machine learning algorithms.

% \subsection{Support Vector Machine}
% Support Vector Machines (SVM) are powerful super- vised learning methods for classification, regression, and outlier detection. In our research, we will deploy SVM to classify companies into predefined categories based on the features extracted from their logos. By training an SVM model on labeled data, we aim to achieve high accuracy in predicting the category of new companies, thereby validating the utility of color and design features in logo categorization.

% \begin{figure}[h]
% \centering
% \includegraphics[width=0.5\textwidth]{images/confusionMatrix.jpg}
% \caption{Confusion Matrix}\label{visina8}
% \end{figure} 

\subsection{Sentiment Analysis}
The Valence Aware Dictionary and Sentiment Reasoner (VADER) is a lexicon and rule-based sentiment analysis tool optimized for social media text. We utilize VADER to analyze the sentiment of Google Maps reviews related to the companies in our dataset. This analysis will help us understand consumers' emotional responses to brands and companies' reputations and complement our examination of logo color impact.

\subsection{Emotion Analysis}
A user often expresses many emotions in a review, not just positive or negative sentiments. These emotions are readily perceived and experienced by other readers. Analyzing these emotions can enhance our understanding of the opinions embedded in reviews \cite{peerj}. Emotion detection in reviews was conducted using \textit{text2emotion 0.0.5} \cite{t2e}. This tool categorizes emotions into five groups: \textit{Happy, Angry, Sad, Surprise,} and \textit{Fear}, and returns a dictionary with scores for each category.

While a high score in a specific emotion could suggest that the review predominantly conveys that emotion, we also consider emotions with moderate and low intensities \cite{peerj}. Therefore, we employ a fuzzy approach to assess the expressiveness of emotions more comprehensively.

Several studies have utilized a fuzzy approach to capture the intensity and classification of emotions, helping to distinguish between various emotional labels \cite{fuzzy_meaning2, Kozlov2023, color_em}. Analysis of emotions can be derived from various channels, such as colors \cite{color_em}, music \cite{music_em}, faces \cite{Kozlov2023}, multichannel \cite{movie}, emoji \cite{Srivastava2021}. Each channel provides unique data for a more detailed understanding of emotional states. We aim to see connections between two channels - text and color.

The main rationale for employing fuzzification in our research is its ability to evaluate emotions in a manner consistent with human judgment. Emotions often lack distinct boundaries and are well-suited to fuzzy set analysis. This method allows for partitioning a range of possible emotions into linguistic categories, providing outputs that are more intuitive and align with human perception. Additionally, fuzzy sets effectively represent the gradual transitions between emotional states \cite{Zadeh1975}. Using linguistic values rather than precise numerical scores aligns more closely with human cognitive processes despite the potential loss in precision \cite{fuzzycw}, \cite{peerj}.

\begin{figure}
    \centering
    \includegraphics[width=\linewidth]{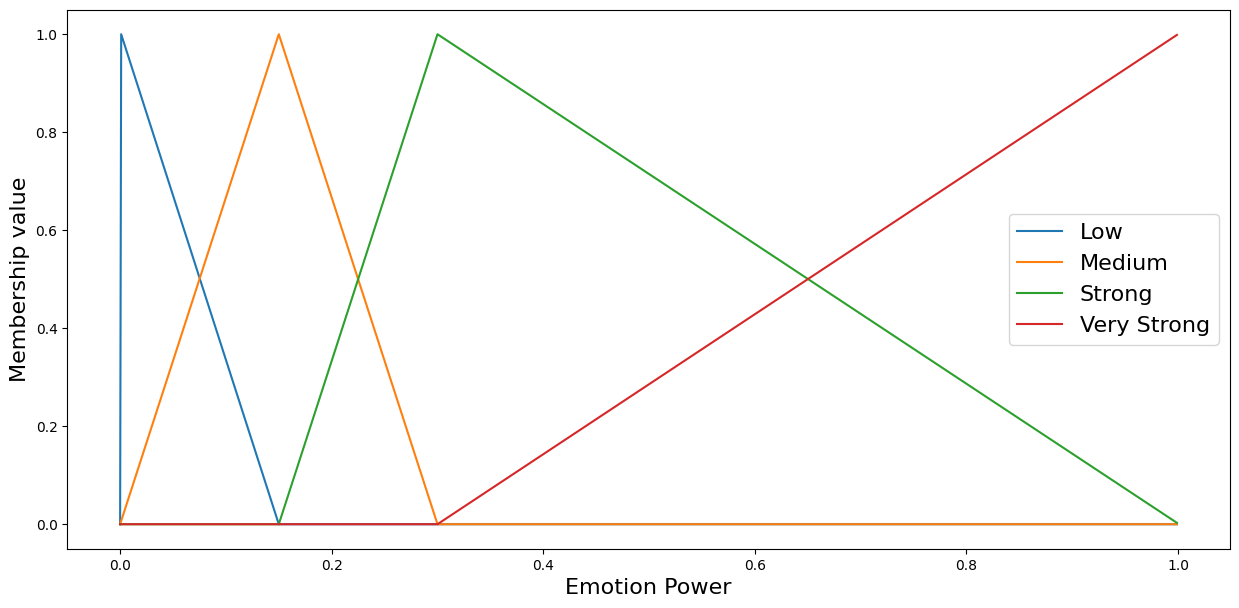}
    \caption{Fuzzification of emotions levels.}
    \label{fig:fuzzy}
\end{figure}

We define \textit{Emotion power} using a series of terms from the fuzzy (linguistic) variable \textit{X = ``Emotion power''}, represented by the primary linguistic terms \textit{L = \{Weak, Medium, Strong, Very Strong\}}, which indicate its intensity level. The accompanying diagram (Fig. \ref{fig:fuzzy}) illustrates the fuzzy sets associated with \textit{Emotion power}. To maintain simplicity, we employ triangular membership functions for each emotion category, which correspond to the linguistic labels that describe the fuzzy variable \textit{Emotion power}. These functions effectively capture the inherent ambiguity in linguistic evaluations. This fuzzy segmentation is based on subjective perception.

As we observe, emotion analysis can assist in identifying more complex feelings embedded in a review. For instance, \textit{`surprise'} is an emotion that can encapsulate both positive and negative sentiments \cite{peerj}. 

% Consider the tweet \textit{``''} Its sentiment score is \textit{Neutral} (-0.043), reflecting a balance of positive and negative words. However, the emotion analysis results show \textit{\{'Happy': 'Medium', 'Sad': 'Weak', 'Fear': 'Weak'\}}. 

\subsection{Color Analysis}

\subsubsection{Color Model}
\begin{figure*}[tb]
\centering
\includegraphics[width=\textwidth]{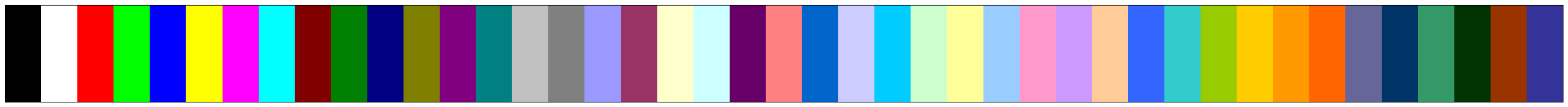}
\caption{Selected colors used in our study}
\label{color_model}
\end{figure*}

Our color model contains 43 items of various saturation, hue, and brightness levels. This selection is specifically suitable for marketing and can vary across contexts. Fig. \ref{color_model} illustrates the selected colors.

\subsubsection{CIEDE2000 Color Difference}
The CIEDE2000 color-difference formula extends earlier CIELAB color-difference formulas, providing a method to measure the perceived difference between two colors \cite{Luo2001}. This formula is particularly useful in applications where precise color matching is crucial, such as textile manufacturing and graphic design. 

The CIEDE2000 color difference formula is given by \cite{Luo2001}:

\begin{equation}
\tiny
\Delta E_{00} = \sqrt{\left(\frac{\Delta L'}{k_L S_L}\right)^2 + \left(\frac{\Delta C'}{k_C S_C}\right)^2 + \left(\frac{\Delta H'}{k_H S_H}\right)^2 + R_T \frac{\Delta C' \Delta H'}{k_C k_H S_C S_H}}
\label{eq:deltaE00}
\end{equation}

where:
\begin{itemize}
    \item $\Delta L'$, $\Delta C'$, and $\Delta H'$ are the differences in lightness, chroma, and hue, respectively, adjusted for the testing conditions.
    \item $S_L$, $S_C$, and $S_H$ are the scaling factors for lightness, chroma, and hue.
    \item $k_L$, $k_C$, and $k_H$ are the parametric weighting factors typically set to 1 under reference conditions.
    \item $R_T$ is the rotation function accounting for the interaction between chroma and hue in the blue region.
\end{itemize}

The components are calculated as follows:
\[
\Delta L' = L_2^* - L_1^* \quad \text{(lightness difference)}
\]
\[
\Delta C' = C_2^* - C_1^* \quad \text{(chroma difference)}
\]
\[
\Delta H' = 2 \sqrt{C_1^* C_2^*} \sin \left(\frac{\Delta h'}{2}\right) \quad \text{(hue difference)}
\]
\[
R_T = -2 \sqrt{\frac{C_1^* C_2^*}{C_1^* + C_2^*}} \sin \left( \delta h \right) \quad \text{(rotational term)}
\]

The differences in hue angle $\Delta h'$ and the rotation term $\delta h$ are particularly complex due to the trigonometric adjustments and are computed based on the average hue.

\subsubsection{K-Means Clustering based Color Palette Extraction}
K-Means clustering, an unsupervised machine learning algorithm \cite{jumb2014color},  \cite{aruzhan}, will be employed to identify dominant colors in the logo images. This method clusters data points (in this case, pixel values) into groups based on similarity. For example, our implementation of K-means clustering on Google’s logo effectively identifies the dominant colors, providing both a visual representation and a list of RGB values for these colors. These values are then used to construct features relevant to our analysis.

\begin{figure}[tb]
\centering
\includegraphics[width=0.5\textwidth]{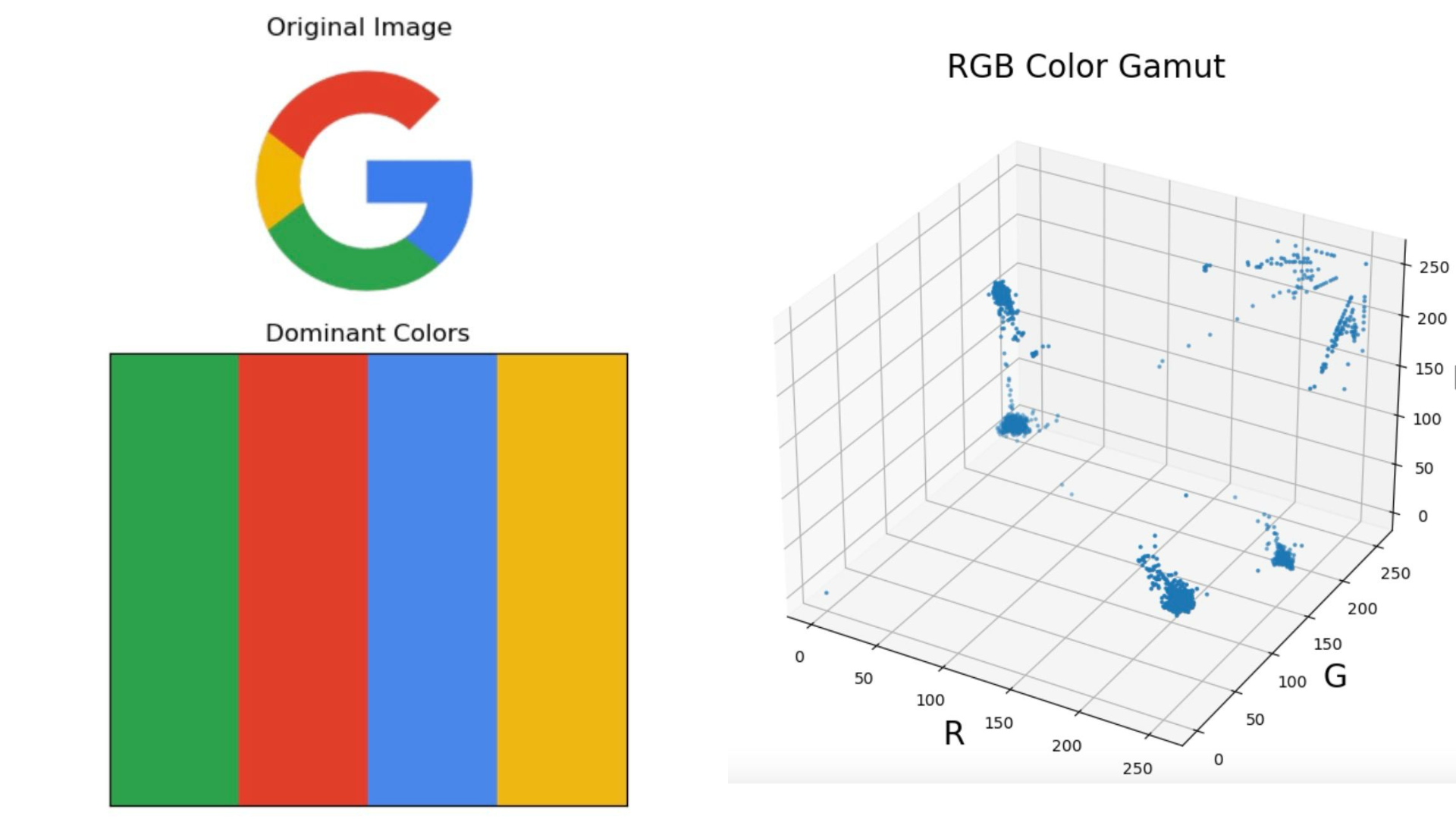}
\caption{Fetching dominant colors from Google's Logo using K-Mean Clustering}
\label{google_ex}
\end{figure}
\begin{figure}[tb]
\centering
\includegraphics[width=0.5\textwidth]{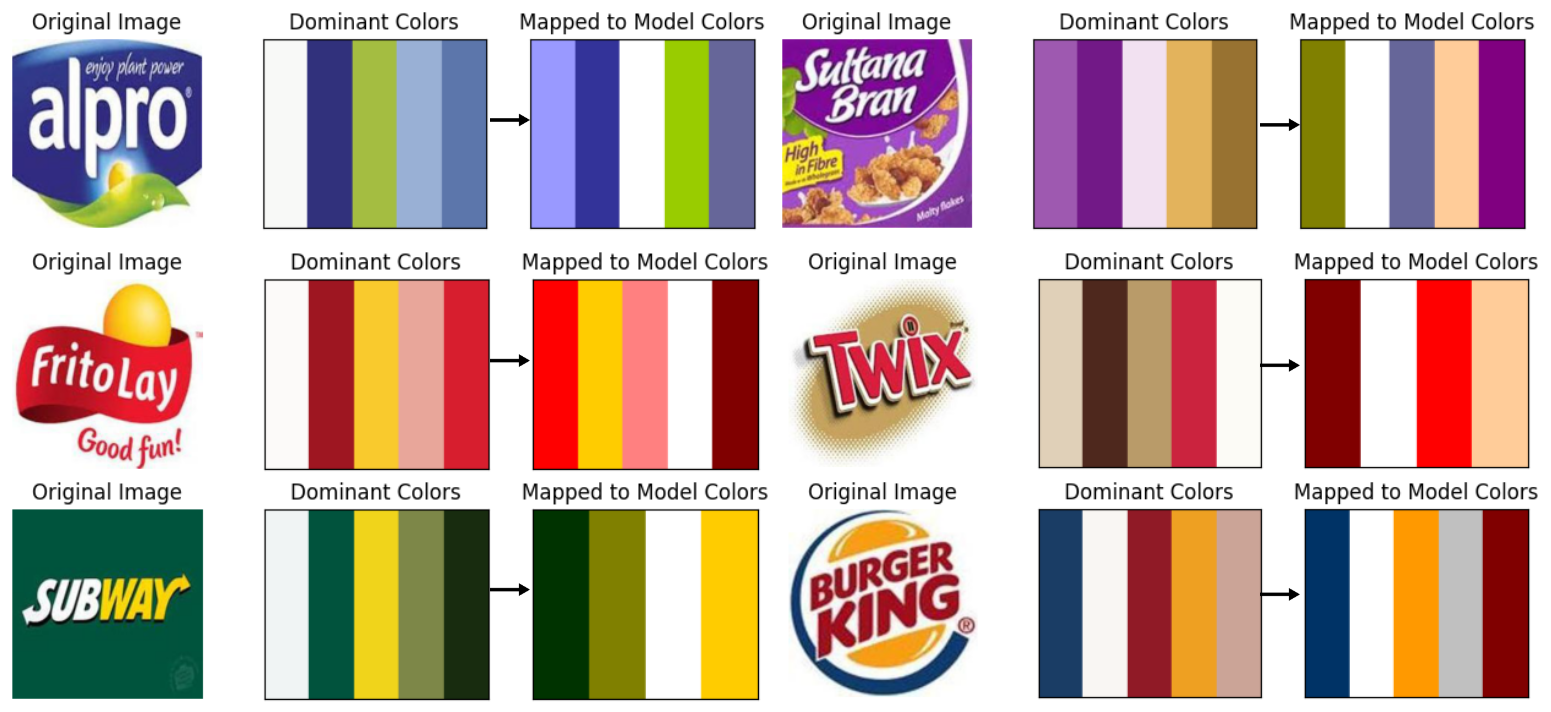}
\caption{Fetching dominant colors from logo images using K-Means Clustering and mapping them to selected color model}
\label{kmeans_ex}
\end{figure}
The extraction of dominant colors from logos is crucial for assessing a brand's visual identity. We use k-means clustering, an effective method for segmenting color data into prominent groups. The methodology is outlined as follows:

\begin{itemize}
    \item \textbf{Preprocessing. } Convert each logo into a set of pixels in a three-dimensional RGB color space.
    
    \item \textbf{Initialization. } Choose $k$, the number of desired clusters corresponding to the number of dominant colors to be identified. Initialize $k$ centroids randomly within the color space of the image.
    
    \item \textbf{Clustering. } Assign each pixel to the nearest centroid based on the Euclidean distance in the RGB space. The distance $d$ between two colors $c_1$ and $c_2$ is calculated as:
 \begin{equation}
 \small
    d(c_1, c_2) = \sqrt{(r_2 - r_1)^2 + (g_2 - g_1)^2 + (b_2 - b_1)^2}
    \label{eq:rgbDistance}
\end{equation}
    where $r$, $g$, and $b$ are the colors red, green, and blue components.
    
    \item \textbf{Centroid Adjustment.} Update each centroid to be the mean of the pixels assigned to it. The mean for each RGB component is recalculated to reflect the new center of the cluster.
    
    \item \textbf{Iteration.} Repeat the clustering and centroid adjustment steps until the centroids no longer change significantly, indicating that the dominant colors have been effectively captured.
\end{itemize}

The result is a set of $k$ centroids, where each centroid represents a dominant color in the logo. These colors serve as a quantitative foundation for subsequent analysis of marketing and consumer perception studies. 

Fig. \ref{google_ex} illustrates the example of applying k-means clustering to fetch the main logo's colors. Fig. \ref{kmeans_ex} shows the example of applying this algorithm to some of the brand logos from the dataset and subsequent mapping to selected colors in our model based on CIEDE difference.

% \begin{figure}[h]
% \centering
% \includegraphics[width=0.5\textwidth]{images/logosAndDominantColors.jpeg}
% \caption{Logo images and Dominant Colors}
% \end{figure}

% \subsection{Regression Analysis}
% Regression analysis provides a robust statistical frame- work to identify and quantify the correlation between company values and reputation. By understanding this rela- tionship, companies can make better strategies to enhance their reputation based on key value drivers. This analysis helps identify the strength and nature of the correlation between these factors.
% By integrating these techniques, our research approach is designed to comprehensively analyze the relationship between logo color and consumer perception, providing actionable insights for brand management and marketing strategies in the food industry.

\section{Results}

\begin{figure}[tb]
\centering
\includegraphics[width=0.5\textwidth]{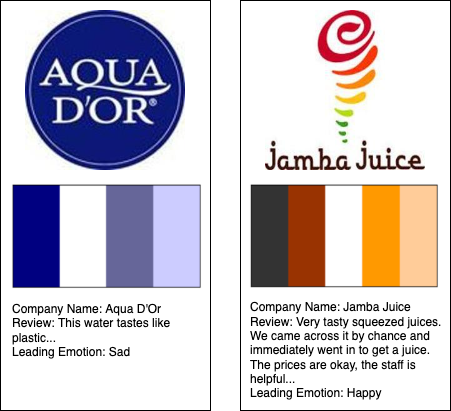}
\caption{Examples of extracted colors, review and leading emotion}
\label{color_review}
\end{figure}

Fig. \ref{color_review} shows examples of company logo images, some reviews and their emotions, and the dominant color palette.

\begin{table}[!h]
    \centering
    
    \begin{tabular}{lr}
    \toprule
    \textbf{Sentiment} & \textbf{Reviews Count} \\
    \midrule
    Positive & 3846 \\
    Neutral & 26110 \\
    Negative & 113 \\
    \midrule
    \textbf{Total}    & 30069 \\
    \bottomrule
    \end{tabular}
    \caption{Distribution of Leading Sentiment}
    \label{tab:sentiment_counts}
\end{table}

Table \ref{tab:sentiment_counts} illustrates the count of reviews categorized by sentiment, revealing a significant prevalence of Neutral reviews at 26,110, followed by Positive reviews numbering 3,846, and a minimal number of Negative reviews at 113. In total, there are 30,069 reviews, indicating that the bulk of the reviews are Neutral, with Positive reviews being considerably fewer, and Negative reviews being rare.

\begin{table}[tb]
    \centering
    
    \begin{tabular}{lr}
    \toprule
    \textbf{Emotion} & \textbf{Companies Count} \\
    \midrule
    Happy    & 399 \\
    Fear     & 215 \\
    Sad      & 23 \\
    Surprise & 7 \\
    Angry    & 0 \\
    \midrule
    \textbf{Total}    & 644 \\
    \bottomrule
    \end{tabular}
    \caption{Distribution of Leading Emotions}
    \label{tab:leading_emotion_names}
\end{table}

Table \ref{tab:leading_emotion_names} displays the frequency of different leading emotions identified in the dataset. As we can see, some emotions are more prevalent than others, with "Happy" being one of the most frequently observed emotions, while there are no companies identified with the leading emotion of "Angry".

\begin{figure*}[h!]
    \centering
    \begin{subfigure}[b]{0.77\textwidth}
        \includegraphics[width=\textwidth]{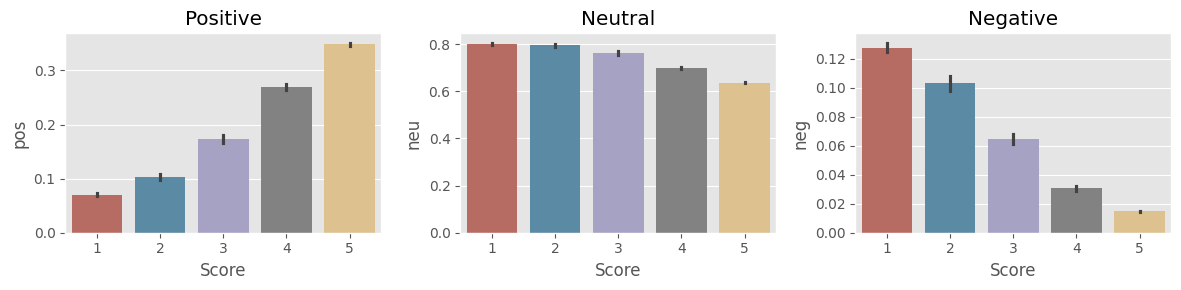}
        \caption{Positive, Neutral and Negative scores}
        \label{fig:sub1}
    \end{subfigure}
    \hfill
    \begin{subfigure}[b]{0.22\textwidth}
        \includegraphics[width=\textwidth]{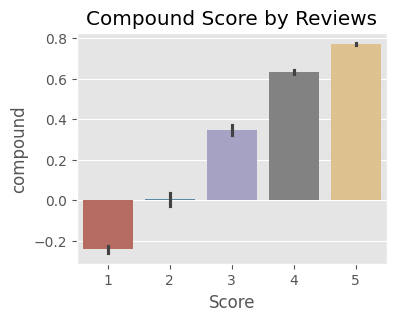}
        \caption{Compound score}
        \label{fig:sub2}
    \end{subfigure}
    \caption{Sentiment Analysis Results}
    \label{fig:test}
\end{figure*}

\begin{figure*}[h]
    \centering
    \includegraphics[width=0.9\textwidth]{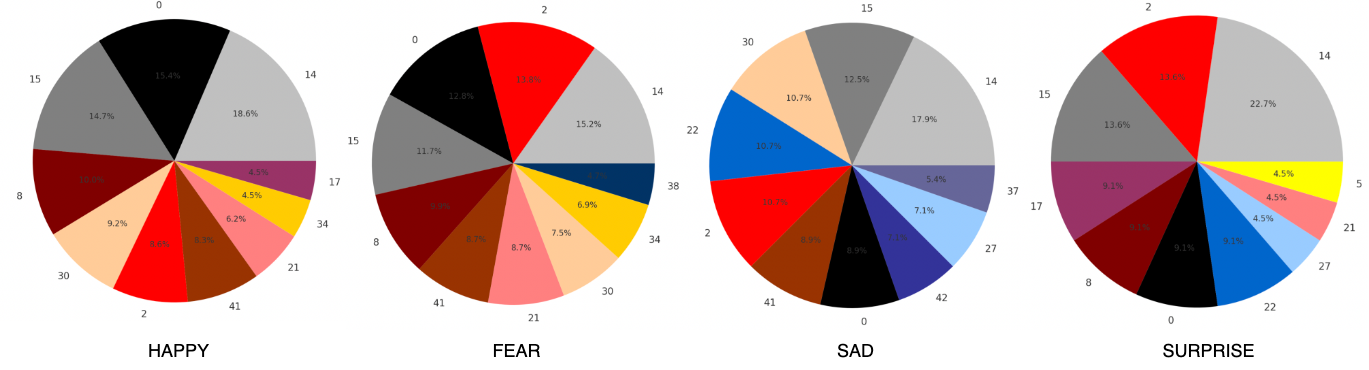}
    \caption{Color-emotion association results in the form of palettes}
    \label{color_result}
\end{figure*}

Fig. \ref{fig:sub1} illustrates that higher review scores (4 and 5) are associated with increased positive sentiment and decreased negative sentiment, indicating more favorable language in these reviews. Lower review scores (1 and 2) show higher negative and slightly elevated neutral sentiments. This pattern demonstrates that higher-rated reviews are more emotionally positive and less neutral or negative.

The analysis of 30,069 reviews reveals a clear positive correlation between review scores and sentiment. Fig. \ref{fig:sub2} shows that reviews with a score of 1 demonstrate predominantly negative sentiment, with an average compound score below zero. As the scores increase, the sentiment becomes progressively more positive, with score 3 showing a moderately positive sentiment, score 4 indicating a stronger positive sentiment, and score 5 demonstrating the most positive sentiment with the highest average compound score. 

Now let us proceed to the main findings of our study. Fig. \ref{color_result} illustrates the pie charts for Happy, Fear, Sad, and Surprise emotions. Across all emotions, Silver (id=14) emerges as a dominant color, while Black (id=0) and Gray (id=15) also frequently appear. Red (id=2) is notably prevalent in all emotions. Happy emotions include lighter, more vibrant colors like light blue and yellow, while Fear also has darker hues like dark blue. Sad emotions are characterized by multiple colors from the blue category. Surprise, with a mix of both light and dark shades, underscores its unpredictable nature. These patterns align with universal color psychology.

Table \ref{tab:color_palettes} illustrates the dominant colors separating the common colors for the food domain. As we can see, while common colors like Silver, Black, and Red span multiple emotions, each emotion's unique palette highlights its distinct psychological impact.

In our analysis, we utilized a color model encompassing 43 distinct colors, including pastel colors, bright colors, highly saturated colors, and dark colors. However, our findings indicate that the dominant colors in the food and beverage industry are predominantly highly saturated, dark, and bright hues. Pastel colors such as beige, light blue, and light pink did not emerge as dominant in this context.

Notably, red, black, and gray emerged as common colors across various emotional spectrums in the food context, similar to recent studies highlighting the dominance of brown in art \cite{color_em}. Our results also reaffirmed certain universal color-emotion associations, such as happiness with yellow, sadness with blue, and surprise with bright colors \cite{color_em}, \cite{Fugate2019}, \cite{Jonauskaite2020}, \cite{Hibadullah2015}.

\begin{table*}[h]
    \centering
    \begin{tabular}{|c|c|c|c|c|c|c|c|c|c|c|}
        \hline
        Emotion & Color 1 & Color 2 & Color 3 & Color 4 & Color 5 & Color 6 & Color 7 & Color 8 & Color 9 & Color 10 \\
        \hline
        Happy & 14 & 2 & 15 & 0 & 8 & 41 & 30 & 21 & 34 & 17 \\
        & \cellcolor[RGB]{192,192,192} & \cellcolor[RGB]{255,0,0} & \cellcolor[RGB]{128,128,128} & \cellcolor[RGB]{0,0,0} & \cellcolor[RGB]{128,0,0} & \cellcolor[RGB]{153,51,0} & \cellcolor[RGB]{255,204,153} & \cellcolor[RGB]{255,128,128} & \cellcolor[RGB]{255,204,0} & \cellcolor[RGB]{153,51,102} \\
        \hline
        Fear & 14 & 2 & 15 & 0 & 8 & 41 & 21 & 30 & 34 & 38 \\
        & \cellcolor[RGB]{192,192,192} & \cellcolor[RGB]{255,0,0} & \cellcolor[RGB]{128,128,128} & \cellcolor[RGB]{0,0,0} & \cellcolor[RGB]{128,0,0} & \cellcolor[RGB]{153,51,0} & \cellcolor[RGB]{255,128,128} & \cellcolor[RGB]{255,204,153} & \cellcolor[RGB]{255,204,0} & \cellcolor[RGB]{0,51,102} \\
        \hline
        Sad & 14 & 2 & 15 & 0 & 41 & 30 & 22 & 42 & 27 & 37 \\
        & \cellcolor[RGB]{192,192,192} & \cellcolor[RGB]{255,0,0} & \cellcolor[RGB]{128,128,128} & \cellcolor[RGB]{0,0,0} & \cellcolor[RGB]{153,51,0} & \cellcolor[RGB]{255,204,153} & \cellcolor[RGB]{0,102,204} & \cellcolor[RGB]{51,51,153} & \cellcolor[RGB]{153,204,255} & \cellcolor[RGB]{102,102,153} \\
        \hline
        Surprise & 14 & 2 & 15 & 0 & 8 & 17 & 22 & 27 & 21 & 5 \\
        & \cellcolor[RGB]{192,192,192} & \cellcolor[RGB]{255,0,0} & \cellcolor[RGB]{128,128,128} & \cellcolor[RGB]{0,0,0} & \cellcolor[RGB]{128,0,0} & \cellcolor[RGB]{153,51,102} & \cellcolor[RGB]{0,102,204} & \cellcolor[RGB]{153,204,255} & \cellcolor[RGB]{255,128,128} & \cellcolor[RGB]{255,255,0} \\
        \hline
    \end{tabular}
    \caption{Top 10 Dominant Colors for Each Emotion}
    \label{tab:color_palettes}
\end{table*}

\section{Limitations and Future Works}
The study has some limitations. Firstly, our dataset comprises 644 companies within the food sector. This domain-specific focus may limit the generalizability of our findings. Next, we focused on the dominant colors extracted from brand logos, which are not the sole influencers of consumer perception.  Thus, focusing on logo colors may not capture the full spectrum of factors affecting consumer emotional responses. Moreover, the emotion extraction process from Google Maps reviews presents potential biases. Due to the variability in individual expression, user-generated content may not consistently reflect genuine emotional states. 

For future work, our current model could be enhanced by incorporating a broader range of colors. Additionally, including multiple data sources, such as social media platforms besides reviews, could provide a more complete view of consumer sentiment. Future research should examine color-emotion associations in other domains, such as accessories, clothing, and cosmetics, to generalize our findings across various industries. Additionally, multimodal data integration (e.g., brand messaging) would allow for a more comprehensive exploration of the connection between various brand elements and consumer sentiment.

\section{Conclusion}

We conducted our study by examining the primary colors in the logos of 644 food and beverage companies using k-means clustering. Then, we linked these colors to emotional reactions derived from Google Maps reviews (n=30,069), using fuzzy logic to sort these emotions by intensity. This allowed us to determine distinct color-emotion pairings.

Our study findings have shown that colors significantly influence how consumers perceive and emotionally respond to food and beverage brands. We found that brands predominantly evoke positive emotions, with Happiness being the most common and no evidence of Anger. This indicates that companies likely choose colors that avoid negative reactions. The frequent use of red and gray in logos across different emotional responses suggests these colors are preferred for their broad appeal. Our findings also confirm specific color-emotion associations, such as yellow with Happiness, blue with Sadness, and bright colors with Surprise. These insights could be invaluable for marketers aiming to evoke desired emotional responses from consumers. enhance brand recognition and loyalty through strategic color choices in logos.

% Future works to do:
% \begin{enumerate}
%     \item DONE Python Script for GoogleMaps Reviews extraction 
%     \item DONE Extend Logos Dataset or Find new one 
%     \item Re-train model on large dataset to get better results
%     \item Combine Visual and Sentiment Analysis to find correlation
%     \item Train Model to recommend color palette for Company category (text-to-color!)
%     \item Apply Fuzzy
% \end{enumerate}

\section*{Acknowledgment}
This research has been funded by the Science Committee of the Ministry of Science and Higher Education of the Republic of Kazakhstan (Grant No. AP22786412)

\bibliography{references}

\end{document}